\documentclass[a4paper,fleqn,usenatbib]{mnras}

\usepackage{newtxtext,newtxmath}

\usepackage[T1]{fontenc}
\usepackage{ae,aecompl}

\usepackage{graphicx}	
\usepackage{amsmath}	
\usepackage{amssymb}	

\title[Forecast on lepton asymmetry from future CMB experiments]{Forecast on lepton asymmetry from future CMB experiments}

\author[A Bonilla, R C Nunes and E M C Abreu]{
Alexander Bonilla,$^{1}$\thanks{E-mail: abonilla@fisica.ufjf.br}
Rafael C Nunes,$^{2}$\thanks{E-mail: rafadcnunes@gmail.com (corresponding author)}
Everton M C Abreu,$^{3,1,4}$\thanks{E-mail: evertonabreu@ufrrj.br}
\\
$^{1}$Departamento de F\'isica, Universidade Federal de Juiz de Fora, 36036-330, Juiz de Fora, MG, Brazil\\
$^{2}$Divis\~ao de Astrof\'isica, Instituto Nacional de Pesquisas Espaciais, Avenida dos Astronautas 1758, S\~ao Jos\'e dos Campos, 12227-010, SP, Brazil\\
$^{3}$Grupo de F\'isica Te\'orica e F\'isica Matem\'atica, Departamento de F\'isica, Universidade Federal Rural do Rio de Janeiro, 23890-971,\\ 
Serop\'edica, RJ, Brazil \\
$^{4}$Programa de P\'os-Gradua\c{c}\~ao Interdisciplinar em F\'isica Aplicada, Instituto de F\'{i}sica, Universidade Federal do Rio de Janeiro-UFRJ, \\ 
21941-972, Rio de Janeiro, RJ, Brazil}

\date{Accepted XXX. Received YYY; in original form ZZZ}

\pubyear{2019}

\begin{document}
\label{firstpage}
\pagerange{\pageref{firstpage}--\pageref{lastpage}}
\maketitle

\begin{abstract}

\noindent We consider a cosmological lepton asymmetry in the form of neutrinos and impose new expected sensitivities on such asymmetry through the 
degeneracy parameter ($\xi_{\nu}$) by using some 
future CMB experiment configurations, such as CORE and CMB-S4. Taking the default scenario with three neutrino states, 
we find $\xi_{\mu} = 0.05 \pm 0.10 \, (\pm \, 0.04)$, from CORE (CMB-S4) at 95 percent CL, 
respectively. Also, within this scenario, we evaluate the neutrino mass scale, obtaining that the normal hierarchy mass scheme is privileged. 
Our results are an update concerning on the cosmological lepton asymmetry and 
the neutrino mass scale within this context, from which can bring a perspective on the null hypothesis for $\xi_{\nu}$ (and its effects 
on $\Delta N_{\rm eff}$), 
where perhaps, \textbf{$\xi_{\nu}$}  may take a non-null value up to 95 percent CL from future experiments such as CMB-S4. 
Sensitivity results for CMB-S4 obtained here not including all expected systematic errors.
\end{abstract}

\begin{keywords}
Cosmic neutrino background -- Observational constraints -- Degeneracy parameter
\end{keywords}



\section{Introduction}

The lepton asymmetry of the Universe, represented by neutrinos and antineutrinos, is nowadays one of the most weakly constrained cosmological parameter.
Although the baryon number asymmetry is well measured
from cosmic microwave background (CMB) constraints concerning the baryon density, the lepton asymmetry could be larger by many orders of magnitude and not of the same order as expected
by the Big Bang Nucleosynthesis (BBN) considerations \citep{2017PhRvD..95d3506B}. The presence of a large lepton asymmetry can be considered as an excess the neutrinos on antineutrinos or vice versa, 
which can be a requirement due to both the charge neutrality of the Universe and it is possibly hidden in the cosmic neutrino background ($C\nu B$) and it can be imprinted on cosmological observation. 
For instance, from CMB anisotropy \citep{Castorina,Dominik02}. The large neutrino asymmetries have consequences in the early Universe phase transitions, cosmological magnetic fields
and dark matter relic density (see \citep{Schwarz,Semikoz,Stuke,2017JCAP...04..048B}, for more details). Other effects due to the asymmetric leptonic can be considered as changes in the decoupling 
temperature of $C\nu B$ \citep{Freese,Kang}, the time equivalence between the energy densities of radiation and matter, the production of primordial light elements at BBN \citep{Sarkar},
an excess in the contribution of the total radiation energy density and the expansion rate of the Universe \citep{Giusarma,2015PhRvD..92l3535A}, photon decoupling \citep{xi7}, among others. These changes 
can affect the evolution of the matter density perturbations in the Universe, which has effects not only on the CMB anisotropies, but also on the formation, evolution and distribution of
the  large-scale structure (LSS) of the Universe \citep{book,2015PhRvD..92l3535A}. The effects of the cosmological neutrinos on both the CMB and LSS are only gravitational, since they are decoupled 
(free streaming particles) 
at the time of recombination and structure formation. The LSS formation is more sensitive to the neutrino masses than CMB. The increasing of the structure is driven by the cosmic expansion and
self-gravity of matter perturbations, both affected by the massive neutrinos. Nevertheless, the relic neutrino slows down the growth of structures due to its high thermal speed, leading 
to a suppression of the total matter power spectrum \citep{Ali}. On the other hand, the gravitational lensing of CMB and the integrated Sachs-Wolf effect are also modified by the presence
of massive neutrinos \citep{Abazajian}. The effect of massive neutrinos in the non-linear growth structure regime has been recently studied by \cite{Zeng}.\\

The neutrinos properties are very important in the determination of the dynamics of the Universe inferring direct effects on cosmological sources 
and consequently in 
the estimation of cosmological parameters (see \citep{Dolgov,Lesgourgues,Abazajian,Wang2,Lorenz,Vagnozzi1,Yang,DiVal2,Li,WangLF,Wang}). 
The parameters that characterize the neutrinos effects on cosmological probes 
are the total neutrino mass $\Sigma m_{\nu}$ and the effective number of species $N_{\rm eff}$. 
Altogether, the up dated constraint on the neutrino mass scale is $\Sigma m_{\nu} < 0.12$ eV at 95 percent C.L.  and $N_{\rm eff}=2.99\pm 0.17$ at 95 percent C.L. 
from the final full-mission Planck measurements of the CMB anisotropies \citep{Planck2018}. 
In the case of the three active neutrino flavors with zero asymmetries and a standard thermal history, 
the value of effective number of species is the well-known $N_{\rm eff} = 3.046$ \citep{book} 
and an improved calculation $N_{\rm eff} = 3.045$ \citep{de Salas}, but the presence of neutrino asymmetries can increase this number without 
the need to introduce new relativistic species. In general terms, any excess over this value can be parameterized through 
$\Delta N_{\rm eff} = N_{\rm eff} - 3.046$, which in principle is assumed to be some excess of the number of relativistic relics degrees of freedom, 
known in the literature as dark radiation
(see \cite{Nunes} for recent constraints on $\Delta N_{\rm eff}$).\\ 

Finally, and more important to our work, is to consider the aforementioned cosmological lepton symmetry, which is another natural extension on the neutrino physics properties. 
This property is usually parameterized by the so-called degeneracy parameter \textbf{$\xi_{\nu} = \mu_{\nu}/T_{\nu 0}$}, where \textbf{$\mu_{\nu}$} is the neutrino chemical potential and $T_{\nu 0}$ is the 
current temperature of the relic neutrino spectrum $T_{\nu 0}\approx1.9K$. 
We can assign to chemical potentials a label of its eigenstates of mass, such that $\lbrace u_i \rbrace$ is for neutrinos and $\lbrace - u_i \rbrace$ for antineutrinos. 
If the neutrinos are Majorana particles, then they must have $u_i = 0$, and if $u_i \neq 0$, neutrinos are Dirac fermions thus, evidence on the null hypothesis is necessary to help to solve this 
question \citep{Mangano}. The difference between $\lbrace \xi_i \rbrace$ and $\lbrace - \xi_i \rbrace$ determines the asymmetry between the density of neutrinos and antineutrinos. 
Then, the presence of a relevant and non-zero $\xi_{\nu}$  have some cosmological implications \citep{xi6,xi10,xi7,xi5,xi4,xi8,xi1,xi9,xi3,xi2,Dominik01,xi11,Dominik02}. 
From the particle physics point of view, to measure the lepton asymmetry of the Universe is crucial to understand some of particle physics processes that might have taken place in early Universe 
at high energies, including the better constraint on models for the creation of matter-antimatter asymmetry in the Universe \citep{Affleck,Casas,Canetti}. The tightest constraints on lepton asymmetry 
at present are commonly based on a combination of CMB data via constraints on the baryon density and measurements of the primordial abundances of light elements \citep{xi2,Mangano,Cooke}.\\ 

In this work, our main target is to obtain new and precise limits on the cosmological lepton asymmetry, measured in terms of the degeneracy parameter $\xi_{\nu}$, as well as 
the neutrino mass scale, taking as a basis, the configurations of future CMB experiments such as CMB-CORE and CMB-S4.\\

This paper is organized as follows. In the next section, we briefly comment on the $C\nu B$ and the cosmological lepton asymmetry. In section \ref{methodology}, 
we present the methodology used to obtain the forecasts from CMB-CORE and CMB-S4 experiments. In section \ref{Results}, we present our results and discussions. Finally, section \ref{conclusions} our final considerations.

\section{Cosmic neutrino background and neutrino asymmetry}
\label{CNB}

The current contribution of neutrinos to the energy density of the Universe is given by,

\begin{eqnarray}
\rho_{\nu} = 10^4 h^2 \Omega_{\nu} eV cm^{-3},
\end{eqnarray}
where \textbf{$\Omega_{\nu}=\rho_{\nu}/\rho_{cri}$} is the neutrino energy density in units of critical density. As usual, relativistic neutrinos contribute to the total energy density of radiation $\rho_{r}$, typically parametrized as

\begin{eqnarray}
\rho_{r} = \left( \rho_{\gamma} + \rho_{\nu} \right) = \left( 1 + \frac{7}{8}\Big(\frac{4}{11} \Big)^{4/3} N_{\rm eff} \right)  \rho_{\gamma},
\end{eqnarray}
where $\rho_{\gamma}$ is the energy density of photons, the factor $7/8$ is due to the neutrinos that are fermions and $N_{\rm eff}=3.046$ the value 
of the effective number of neutrinos species in the standard case, 
with zero asymmetries and no extra relativistic degrees of freedom. 
Neutrinos become nonrelativistic when their average momentum falls below their mass. 
At the very early Universe, neutrinos and antineutrinos 
of each flavor $\nu_i$ ($i = e,\mu,\tau$) behave like relativistic particles. 
Both the energy density and pressure of one species of massive degenerate neutrinos and antineutrinos are 
described by (ley us use here the unit system where $\hbar = c = k_B = 1$)

\begin{eqnarray}
 \rho_{\nu_i} +  \rho_{\bar{\nu_i}} = T^4_{\nu} \int \frac{d^3q}{2(\pi)^3}  E_{\nu_i} (f_{\nu_i}(q) + f_{\bar{\nu_i}}(q)) 
\end{eqnarray}
and 

\begin{eqnarray}
3 (p_{\nu_i} +  p_{\bar{\nu_i}}) = T^4_{\nu} \int \frac{d^3q}{2(\pi)^3} \frac{q^2}{E_{\nu_i}}(f_{\nu_i}(q) + f_{\bar{\nu_i}}(q)),
\end{eqnarray}
where $E^2_{\nu_i} = q^2 + a^2 m_{\nu_i}$ is one flavor neutrino/antineutrino energy and $q = a p$ is the comoving momentum.
The functions $f_{\nu_i}$, $f_{\bar{\nu_i}}$ are the Fermi-Dirac phase space distributions given by

\begin{eqnarray}
f_{\nu_i}(q) = \frac{1}{e^{E_{\nu_i}/T_{\nu} - \xi_{\nu}} + 1}, f_{\bar{\nu_i}}(q) = \frac{1}{e^{E_{\bar{\nu_i}}/T_{\nu} + \xi_{\bar{\nu}}} + 1},
\end{eqnarray}
where \textbf{$\xi_{\nu} = \mu_{\nu}/T_{\nu0}$} is the neutrino degeneracy parameter and $\mu_{\nu}$ is the neutrino chemical potential. At the early Universe, we assumed that neutrinos-antineutrinos are produced in 
thermal and chemical equilibrium. Their equilibrium distribution functions has been frozen from the time of decoupling to the present. Then, as the chemical potential \textbf{$\mu_{\nu}$} scales as $T_{\nu}$, the degeneracy parameter $\xi_{\nu}$ remains constant and it is different from zero if a neutrino-antineutrino asymmetry has been produced before the decoupled. The energy of neutrinos changes according to cosmological redshift 
after decoupling, which is a moment when they are still relativistic. The neutrino degeneracy parameter is conserved and the presence of a significant and non-null $\xi_{\nu}$ have some cosmological implication 
through the evolution of the universe, such as BBN, photon decoupling and LSS, among others 
\citep*[see][]{xi6,xi10,xi7,xi5,xi4,xi8,xi1,xi9,xi3,xi2,xi11}. If $\xi_{\nu}$ remains constant, finite and non-zero after decoupling, then it could lead to an asymmetry on the neutrinos and antineutrinos given by

\begin{eqnarray}
\label{asymmetry}
\eta_{\nu}  \equiv \frac{n_{\nu_{i}}-n_{\bar{\nu}_{i}}}{n_{\gamma}} = \frac{1}{12 \zeta (3)}\sum_{i} y_{\nu 0} \left( \pi^2 \xi_{\nu_i} + \xi_{\nu_i}^3 \right),
\end{eqnarray}
where $n_{\nu_{i}} (n_{\bar{\nu}_{i}})$ is the neutrino (antineutrino) number density, $n_{\gamma}$ is the photon number density, $\zeta (3) \approx 1.20206$,  and $y_{\nu 0}^{1/3}=T_{\nu 0}/T_{\gamma 0}$ is the ratio of neutrino and photons temperature to the present, where $T_{\gamma 0}$ is the temperature of the CMB ($T_{\gamma 0} = 2.726K$).\\

\noindent As we have mentioned above, the neutrino asymmetry can produce changes in the expansion rate of the Universe at early times, which can be expressed as an excess in $N_{\rm eff}$ in the form

\begin{eqnarray}
\label{Delta_Neff}
\Delta N_{\rm eff} = \frac{15}{7} \sum_i \left[ 2 \left( \frac{\xi_{\nu_i}}{\pi} \right)^2 + \left( \frac{\xi_{\nu_i}}{\pi}\right)^4  \right].
\end{eqnarray}
In what follows, let us impose expected sensitivities on $\xi$ by  taking predictions from some future CMB experiments.

\section{Methodology}
\label{methodology}

Let us predict the ability of future CMB experiments to constrain the neutrino lepton asymmetry as well as the neutrino mass scale.
We follow the common approach already used \citep*[see e.g][]{DiVal,Finelli}, on mock data for some possible future experimental 
configurations, assuming a fiducial flat $\Lambda$CDM model compatible with the 
Planck 2018 results. We  have used  the  publicly  available  Boltzmann code CLASS \citep{class} to compute the theoretical CMB angular power spectra 
$C_l^{TT}$, $C_l^{TE}$, $C_l^{EE}$ for temperature, cross temperature-polarization and polarization. Together with the primary anisotropy signal, 
we have also taken into account informations from CMB weak lensing,
considering the power spectrum of the CMB lensing potential $C_l^{PP}$. The missions BB are clearly sensitive also to the BB lensing polarization signal,
but here we take the conservative approach to not include it in the forecasts.\\ 

In our simulations, we have used an instrumental noise given by the usual expression
\begin{eqnarray}
 N_l = w^{-1} \exp(l(l+1)\theta^2/8 \ln(2)),
\end{eqnarray}
where $\theta$ is the experimental FWHM angular resolution, $w^{-1}$ is the experimental power noise expressed in
$\mu$K-arcmin. The total variance of the multipoles $a_{lm}$ is therefore given by the sum of the fiducial $C'_{l}$s with the instrumental noise $N_{l}$.\\

The simulated experimental data are then compared with a theoretical model assuming a Gaussian likelihood $\mathcal{L}$ given by

\begin{eqnarray}
- 2 \ln \mathcal{L} = \sum_l (2l + 1) f_{sky} \Big( \frac{D}{|\bar{C}|} + \ln \frac{|\bar{C}|}{|\hat{C}|} -3 \Big),
\end{eqnarray}
where $\bar{C_l}$ and $\hat{C_l}$ are the assumed fiducial and theoretical spectra plus noise, and $|\bar{C}|$ and $|\hat{C}|$
are the determinants of the theoretical and observed data covariance matrices given by

\begin{eqnarray}
 |\bar{C}| = \bar{C_l}^{TT} \bar{C_l}^{EE}\bar{C_l}^{PP} - (\bar{C_l}^{TE})^2 \bar{C_l}^{PP} - (\bar{C_l}^{TP})^2 \bar{C_l}^{EE},
\end{eqnarray}

\begin{eqnarray}
 |\hat{C}| = \hat{C}^{TT} \hat{C}^{EE} \hat{C}^{PP} - (\hat{C}^{TE})^2 \hat{C}^{PP} - (\hat{C}^{TP})^2 \hat{C}^{EE},     
\end{eqnarray}
$D$ is defined as

\begin{eqnarray}
 D = \hat{C}^{TT} \bar{C_l}^{EE} \bar{C_l}^{PP} + \bar{C_l}^{TT}\hat{C}^{EE}\bar{C_l}^{PP} + \bar{C_l}^{TT}\bar{C_l}^{EE}\hat{C}^{PP}  \nonumber \\  
 - \bar{C_l}^{TE}( \bar{C_l}^{TE} \hat{C}^{PP} + 2 \hat{C}^{TE} \bar{C_l}^{PP})  \nonumber \\ 
 - \bar{C_l}^{TP} (\bar{C_l}^{TP} \hat{C}^{EE} + 2 \hat{C}^{TP} \bar{C_l}^{EE}),
\end{eqnarray}
and finally $f_{sky}$ is the sky fraction sampled by the experiment after foregrounds removal.\\

\noindent In Table \ref{tab1} we have summarized the experimental specifications for CMB-CORE and CMB-S4 data. 
Forecast based on future CMB experiments to probe neutrinos properties were also investigated in
\cite{Capparelli,Brinckmann,Mishra}. Specifically for CMB-S4, we use an experimental specification different from that presented in 
\cite{Mishra} and \cite{S4} ($l_{\rm max} = 5000$), where we avoid systematic erros at highest angular resolution ($l>3000$) 
produced by noise due to information from CMB weak lensing, which could produce more optimistic results than in previous analyzes.

\begin{table*}
\centering
 \caption{ Experimental specifications for CORE and S4 with beamwidth, power noise sensitivities of the temperature and polarization. }
         \label{tab1}
          \begin{tabular}{lccccr}
          \hline
           Experiment               &  Beam    &  Power noise [$\mu$K-arcmin] &  $l_{\rm min}$  & $l_{\rm max}$  & $f_{\rm sky}$ \\
          \hline
          CMB-CORE                      & 6.0      &  2.5                         & 2                & 3000          & 0.7   \\
          CMB-S4                        & 3.0      &  1.0                        & 50               & 3000          &  0.4    \\
          \hline
          \end{tabular}
\end{table*}

\section{Results}
\label{Results}

We have used the public and available CLASS \citep{class} and Monte Python \citep{monte} codes concerning the model considered in the present work, where we introduced the $\xi_{\nu}$ corrections on $N_{\rm eff}$ defined in equation (\ref{Delta_Neff}) in CLASS code. We have considered one massive and two massless neutrino states, as standard in the literature, and we fixed the mass ordering to the normal hierarchy with the minimum masses $\sum m_{\nu} = 0.06$ eV and the expected sensitivities obtained on the total neutrino mass are essentially independent of neutrino mixing parameters how is it concluded in \cite{Castorina}.\\ 

\noindent The individual values of neutrino flavour asymmetries in principle can be different if we take into account the effect of the oscillations and collisions around the epoch of neutrino decoupling, which means that the equations (\ref{asymmetry}) and (\ref{Delta_Neff}) are not necessarily valid \citep{Pastor,Castorina}. However, following \cite{Dominik02} we assume a single value of $\xi_{\nu}$, which means that for the values of the neutrino mixing parameters preferred by global fits of oscillations, and in particular that of $sin^2 \theta_{13}$, the impact of a lepton asymemtry can be approximated by choosing a common value $\xi_{\nu}$ for degenerancy parameters \citep{2002NuPhB.632..363D,xi4,Mangano}.\\

\noindent On the other hand, in \cite{Castorina} it is shown how the addition of flavor oscillations produces strong constraints on the total neutrino asymmetry, whose boundaries are dominated mainly by the limits imposed by BBN tests. However, although the impact that produces the combined effect of BBN and flavor oscillations on these limits is evident, we are more interested in showing the impact of the improvement in the sensitivity of future CMB experiments.\\

\noindent In our forecasts, we have assumed the set of the cosmological parameters: 
\[  
\{100 \omega_{\rm b}, \, \omega_{\rm cdm}, \, \ln10^{10}A_{s}, \,  
 n_s, \, \tau_{\rm reio}, \, H_0, \, \sum m_{\nu}, \, \xi_{\nu} \}.
\]
where the parameters are: baryon density, cold dark matter density, amplitude
and slope of the primordial spectrum of metric fluctuations, optical depth to reionization, Hubble constant, 
neutrino mass scale, and the degeneracy parameter characterizing the degree of leptonic asymmetry, respectively. 
In the forecast, we assume fiducial values of \{ 2.22, 0.119, 3.07, 0.962, 0.05, 68.0, 0.06, 0.05
\footnote{This value is in accordance with the results obtained in \citep{Nunes} (Table III) 
and \cite{Castorina}.}\}, which are assumed from our analysis performed for Planck 2018.\\ 

\begin{table*}
\centering
\caption{Summary of the observational constraints from both CORE and S4 experiments. The notation $\sigma{\rm (CORE)}$ and $\sigma{\rm (S4)}$, 
represents the 68 percent CL estimation on the fiducial values from
CORE and S4, respectively. The parameter $H_0$ is in  km s${}^{-1}$ Mpc${}^{-1}$ units and $\sum m_{\nu}$ is in eV units. }
\label{results}
\begin{tabular} { l l l l l l }
\hline
  Parameter            &  Fiducial value &  $\sigma{\rm (CORE)}$ & $\sigma{\rm (S4)}$    \\
\hline

{$10^{2}\omega_{b }$} & 2.22     & 0.000057   & 0.00012   \\\\

{$\omega_{cdm }  $}    & 0.11919 & 0.00037   & 0.0000093    \\\\

{$H_0$}                & 68.0    & 0.32       & 0.0088    \\\\

{$\ln10^{10}A_{s}$}    & 3.0753  & 0.0056    & 0.0035  \\\\

{$n_{s}         $}     & 0.96229 & 0.0022     & 0.0054   \\\\

{$\tau_{\rm reio}   $} & 0.055   & 0.0028    & 0.00025    \\\\

{$\sum m_{\rm \nu}$}   & 0.06    & 0.024     & 0.00053  \\\\

{$\xi_{\nu}$}          & 0.05    & 0.071     & 0.027  \\\\

\hline
\end{tabular}
\end{table*}

\begin{figure}
   	\includegraphics[width=9cm]{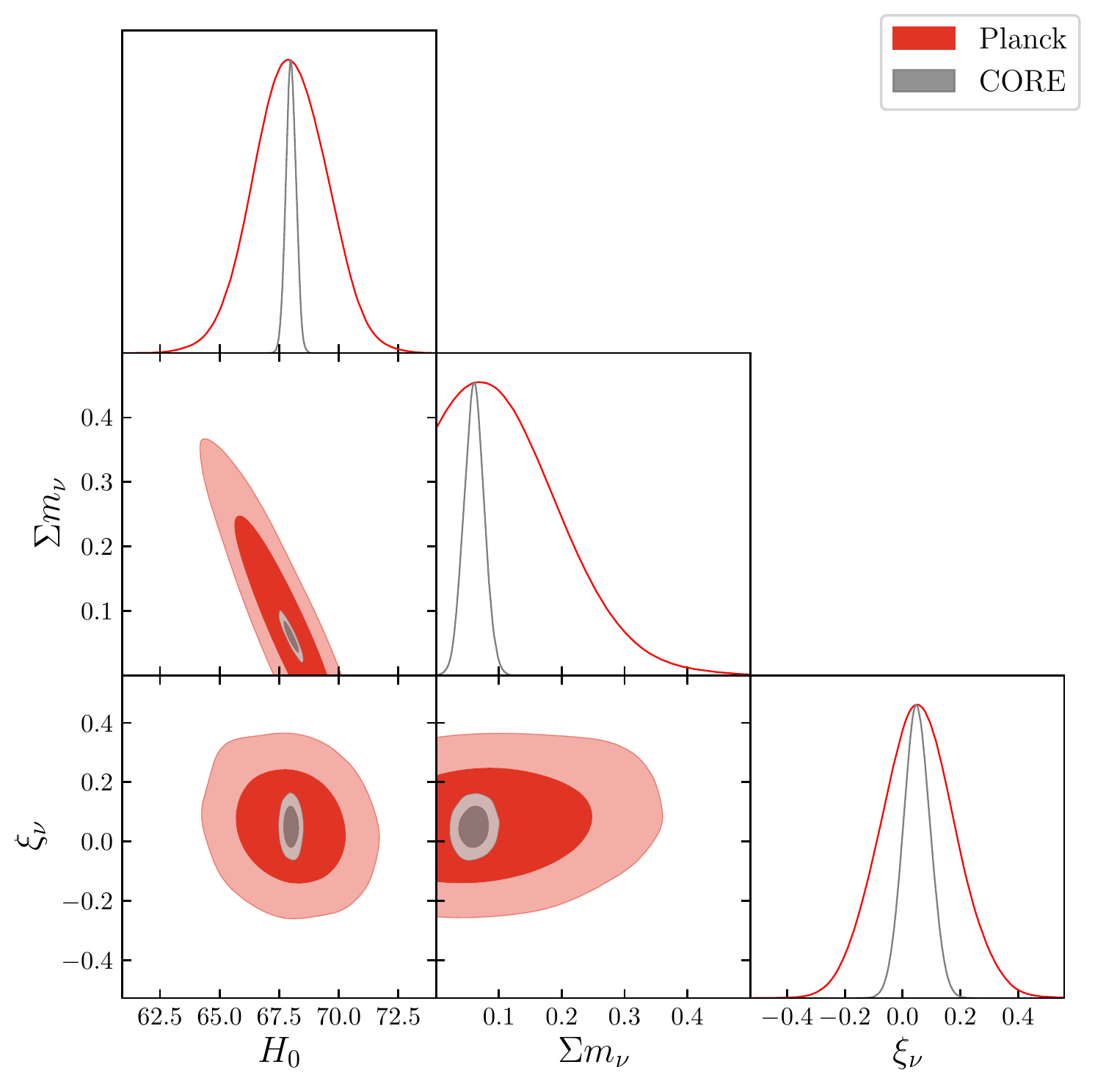}
   	\caption{One-dimensional marginalized distribution and 68 percent CL and 95 percent CL regions for some selected parameters taking into 
   	account Planck and CORE experiments.}
   	\label{Planck_Core} 
\end{figure}

\begin{figure}
   	\includegraphics[width=9cm]{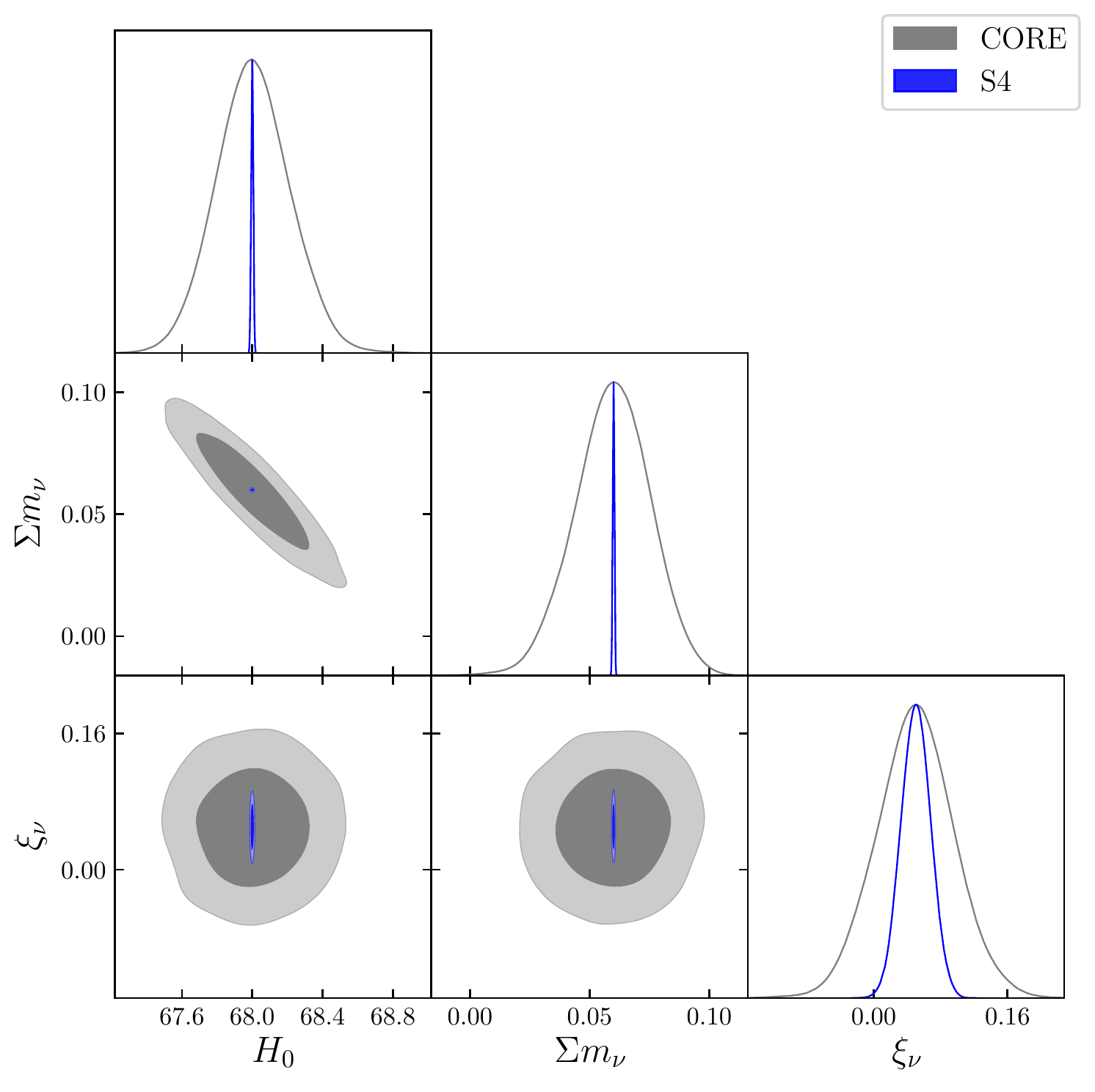}
   	\caption{One-dimensional marginalized distribution and 68 percent CL and 95 percent CL regions for some selected parameters taking 
   	into account CORE and S4 experiments.}
   	\label{Core_S4}
\end{figure}

\noindent Table \ref{results} shows the constraints on the model baseline imposed by the CORE and S4 experiments. Figs \ref{Planck_Core} and \ref{Core_S4} shows the parametric space for some parameters of 
interest in our work, from Planck/CORE and CORE/S4 constraints, respectively. From Planck data, we can note that the degeneracy parameter is constrained to $\xi_{\nu} = 0.05 \pm 0.20$ ($\pm 0.33$) at 68 percent CL and 95 percent CL., which is a result compatible with the null hypothesis even to 1$\sigma$ CL. In \cite{Dominik02}, the authors obtain $\xi_{\nu} = -0.002^{+0.114}_{-0.11}$ at 95 percent CL from Planck data. Evidence for cosmological lepton asymmetry from CMB data have been found by \cite{xi11}.\\ 

\noindent On the other hand, the constraints on the degeneracy parameter are close to the null value also within the accuracy achieved by CORE data, $\xi_{\nu}=0.05\pm 0.071$ ($\pm 0.11$) at 68 percent CL and 95 percent CL, being compatible with the null hypothesis even to 1$\sigma$ CL, as in the case of Planck data, used in this present work. However, with respect to the accuracy obtained by CMB-S4, we find $\xi_{\nu}=0.05\pm 0.027$ ($\pm 0.043$) at 68 percent CL (95 percent CL), respectively. These constraints can rule out the null hypothesis up to 2$\sigma$ CL on $\xi_{\nu}$. In principle this last result can opens the door to the possibility to unveil the physical nature of neutrinos, that is, the neutrinos can be Dirac particles against the null hypothesis and no Majorana particles as established by such hypothesis. However, these results must be firmly established from the point of view of particle physics for example from ground-based experiments such as PandaX-III (Particle And Astrophysical Xenon Experiment III), which to explore the nature of neutrinos, including physical properties such as the  absolute  scale  of  the neutrino masses and the aforementioned violation of leptonic number conservation through Neutrinoless Double Beta Decay (NLDBD)\footnote{In nuclear physics Double Beta Decay (DBD) it's a type of radioactive decay process of second-order weak interactions observed experimentally in several isotopes, in which two electrons (positrons) and 
two antineutrinos (neutrinos) are emitted simultaneously from decaying nucleus (protons into neutrons or vice versa). If neutrinos are Majorana particles, a second DBD mode is possible, where a nucleus can decay again by emitting just two electrons (positrons) without antineutrinos (neutrinos), which are exchanged in the decay of nucleons. } and whose observation will be a clear signal that the neutrinos are their own antiparticles \citep*[for more details see][]{Chen}. These results could be available within the first 5 - 10 years of the next decade.\\

\noindent In Fig. \ref{Planck_Core} we can note that there is a high anti-correlation between the neutrinos' masses and $H_0$, that will increase the tension between the local and global measures of $H_0$, if the masses of the neutrinos increase (therefore they will decrease the value of $H_0$), such that, constraints on those parameters must be cautiously interpreted until such tension can be better understood. Within the standard base-$\Lambda$CDM cosmology, the Planck Collaboration \citep{Planck2018} report $H_0=67.36\pm 0.54\,km\,s^{-1}Mpc^{-1}$  which is about 99 percent away from the locally measured value $H_0=72.24\pm 1.74\,km\,s^{-1}Mpc^{-1}$ reported in \cite{riess}. We obtain, $H_0 = 68.00 \pm 2.32$ ($\pm 3.78$) $km\,s^{-1}Mpc^{-1}$ at 68 percent CL and 95 percent CL, for our model with Planck data, which at least $2\sigma$ can reduce the tension between the global and local value of $H_0$. The difference with our results regarding Planck 2018 is due to our extended parameter space. On the other hand, from the Planck data analysis we can note that the neutrino mass scale is constrained to $\sum m_{\rm \nu} < 0. 36 $ eV at 95 percent CL, 
which is in good agreement with the one obtained by Planck Collaboration, i.e., $\sum m_{\rm \nu} < 0.24 $ eV \citep{Planck2018}. From the $\sum m_{\rm \nu}-H_0$ plane, we note that no relevant changes are obtained with respect to the mass splitting, which requires that $\sum m_{\rm \nu} < 0.1 $ eV to rule out the inverted mass hierarchy ($m_2\gtrsim m_1 \ggg m_3$). However, these results starts to favor the scheme h
of normal hierarchy ($m_1 \ll m_2 < m_3$) which will be evidenced with our results take CORE and S4 predictions. The results from CORE and S4 present considerable improvements with respect to Planck data, see Figs \ref{Planck_Core} and \ref{Core_S4}. With respect to neutrino mass scale bounds imposed from CORE and S4 data, we find the limit $ 0.021 < \sum m_{\rm \nu} \lesssim 0.1$ eV and $0.05913 < \sum m_{\rm \nu} \lesssim 0.061$ eV at 95 percent CL, for CORE and S4, respectively. Thus, unfavorable to inverted hierarchy scheme mass at least at 95 percent CL in both cases.\\

\noindent In the standard scenario of three active neutrinos and if we consider effects of non-instantaneous decoupling, we have $N_{\rm eff}=3.046$, where it is important to make clear that in all the analysis we considered this value as a fixed one. It is well known that the impact of the leptonic asymmetry increase the radiation energy density with the form, $N_{\rm eff} = 3.046 + \Delta N^{\xi_{\nu}}_{\rm eff}$, where $\Delta N^{\xi_{\nu}}_{\rm eff}$ is due to the leptonic asymmetry induced via equation (\ref{Delta_Neff}):

\begin{eqnarray}
\label{Delta_Neff_2}
\Delta N^{\xi_{\nu}}_{\rm eff} = \frac{60}{7} \left( \frac{\xi_{\nu}}{\pi} \right)^2 +  \frac{30}{7}\left( \frac{\xi_{\nu}}{\pi}\right)^4,
\end{eqnarray}
where the sum is $i = 1, 2$ only (two massless neutrino states). Without losing of generality, we can evaluate the contribution $\Delta N^{\xi_{\nu}}_{\rm eff}$ via the standard error propagation theory. We note that, $\Delta N^{\xi_{\nu}}_{\rm eff}=0.002\pm 0.019$ ($\pm0.030$) for Planck data, $\Delta N^{\xi_{\nu}}_{\rm eff} = 0.0022 \pm 0.0083$ ($\pm0.013$) for CORE data and $\Delta N^{\xi_{\nu}}_{\rm eff} = 0.0022\pm 0.0045$ ($\pm0.0059$) for S4 data, all limits at 68 percent and 95 percent CL. Therefore, in general lines, we can assert that the contribution from $\xi_{\nu}$ on $N_{\rm eff}$ are very small. But in the case of CMB-S4, even this contribution being very small, it can be non-null.

\section{Conclusions}
\label{conclusions}

In this work, we have derived new constraints relative to the lepton asymmetry through the degeneracy parameter by using the CMB angular power spectrum from the Planck data and future CMB experiments like CORE 
and CMB-S4. We have analyzed the impact of a lepton asymmetry on $N_{\rm eff}$ where, as expected, we noticed the existence of very small corrections on $\Delta N_{\rm eff}$, but corrections that can not negligible at the level of CMB-S4 experiments, although should be taken into account that sensitivity results obtained from CMB-S4 not including all expected systematic errors, as was previously mentioned \citep*[for similar considerations and more detailed information see][]{2015PhRvD..92l3535A}. Within this cosmological scenario, we have also investigated the neutrino mass scale in combination with the cosmological lepton asymmetry. We have found strong limits on $\sum m_{\rm \nu}$, where the mass scale for both, CORE and CMB-S4 configurations, are well bound to be $\sum m_{\rm \nu} < 0.1$ eV at 95 percent CL, 
therefore, favoring a normal hierarchy scheme within the perspective adopted here.\\ 

\noindent As future perspective, it can be interesting to consider a neutrino asymmetry interaction with the dark sector of the Universe, and to see how this coupling can affect the neutrino and dark matter/dark energy properties, as well as to bring possible new corrections on $\Delta N_{\rm eff}$ due to such interaction, including properly the effect of flavor oscillations and galaxy bias due to neutrinos, which have been recently targeted in literature. For more details about these last topics and a deeper discussion we cordially invite the reader to \cite{Pastor,Castorina}, where can be perceived as a proper implementation of the oscillations can significantly improve the constraints on the main physical properties of massive neutrinos and about the neutrino bias galaxy such as in \cite{Vagnozzi2, Giusarma2}, where is suggested that the proper modeling of the bias parameter is necessary in order to reduce the impact of non-linearities 
and minimization of systematics, in addition to the need to correct for the neutrino-induced scale-dependent bias, whose correct implementation is still under construction within the community of cosmologists.

\section{Acknowledgments}

The authors thank the referee for his/her valuable comments and suggestions. E.M.C. Abreu  thanks CNPq (Conselho Nacional de Desenvolvimento Cient\'ifico e Tecnol\'ogico), Brazilian scientific support federal agency, 
for partial financial support, grant number 302155/2015-5.









%
%


\bsp	
\label{lastpage}
\end{document}